\documentclass[prb,floats,aps,amssymb,preprint,epsf,epsfig,rotate,superscriptaddress]{revtex4}
\usepackage{bm}
\usepackage{graphicx}

\begin{document}
\title{Position-dependent exact-exchange energy for slabs and semi-infinite jellium}

\author{C. M. Horowitz}
\affiliation{Donostia International Physics Center (DIPC), E-20018 Donostia, Basque Country, Spain}
\affiliation{Instituto de Investigaciones Fisicoqu\'imicas Te\'oricas 
y Aplicadas, (INIFTA), UNLP, CCT La Plata-CONICET, Sucursal 4, Casilla de Correo 16, (1900)
La Plata. Argentina}
\author{L. A. Constantin}
\affiliation{Department of Physics and Quantum Theory Group, Tulane University, New Orleans, LA 70118}
\author{C. R. Proetto}
\altaffiliation[Permanent address: ]{Centro At\'omico Bariloche and Instituto Balseiro, 8400, S. C. de Bariloche, 
R\'{i}o Negro, Argentina}
\affiliation{Institut f\"ur Teoretische Physik, Freie Universit\"at Berlin, 
Arnimallee 14, D-14195 Berlin, Germany}
\affiliation{European Theoretical Spectroscopy Facility (ETSF)}
\author{J. M. Pitarke}
\affiliation{CIC nanoGUNE Consolider, Tolosa Hiribidea 76, E-20018 Donostia, Basque Country, Spain}
\affiliation{Materia Kondentsatuaren Fisika Saila and Centro F\'isica Materiales CSIC-UPV/EHU, 644 Posta Kutxatila, E-48080 Bilbo, Basque Country, Spain}

\date\today

\begin{abstract}
The position-dependent exact-exchange energy per particle $\varepsilon_x(z)$ 
(defined as the interaction between a given electron at $z$ and its exact-exchange hole) at metal surfaces 
is investigated,
by using either jellium slabs or the semi-infinite (SI) jellium model. 
For jellium slabs, we prove analytically and
numerically that in the vacuum region far away from the surface
$\varepsilon_{x}^{\text{Slab}}(z \rightarrow \infty) \rightarrow - \; e^{2}/2z$,
{\it independent} of the bulk electron density, which is exactly half the corresponding
exact-exchange potential
$V_{x}(z \rightarrow \infty) \rightarrow - \; e^2/z$ 
[Phys. Rev. Lett. {\bf 97}, 026802 (2006)] of density-functional theory, as occurs in the case of finite systems.
The fitting of $\varepsilon_{x}^{\text{Slab}}(z)$
to a physically motivated image-like expression is feasible, but the resulting 
location of the image plane shows strong finite-size oscillations every time a slab
discrete energy level becomes occupied. For a semi-infinite jellium, the asymptotic behavior of 
$\varepsilon_{x}^{\text{SI}}(z)$ is somehow different. As in the case of jellium slabs
$\varepsilon_{x}^{\text{SI}}(z \rightarrow \infty)$
has an image-like behavior of the form $\propto - \; e^2/z$, but now with a density-dependent coefficient that in general differs from the slab universal coefficient $1/2$. 
Our numerical estimates for this coefficient agree with two previous
analytical estimates for the same. For an arbitrary finite thickness of a jellium slab, we find that the asymptotic limits of
$\varepsilon_{x}^{\text{Slab}}(z)$ and
$\varepsilon_{x}^{\text{SI}}(z)$ only coincide in the low-density limit 
($r_s \rightarrow \infty$),
where the density-dependent coefficient of the semi-infinite jellium approaches the
slab {\it universal} coefficient 1/2.
\end{abstract}

%\pacs{73.20 Dx, 71.35.-y,77.55.+f}

\maketitle

\section{Introduction}

The jellium model of a metal surface, introduced by Bardeen in 1936,~\cite{bardeen}
is the simplest model which reproduces qualitatively, and sometimes quantitatively,
the physical properties of real metal surfaces.~\cite{lang} While in his work Bardeen
applied an approximated Hartree-Fock (HF) theory for the study of the electronic structure,
since the seminal work of Lang and Kohn (LK)~\cite{langkohn} the standard theoretical tool applied
to the study of the electronic structure of metal surfaces has been Density-Functional Theory
(DFT).~\cite{dreizler} As in the original work of Lang and Kohn, most of the subsequent investigations
have applied the Local-Density Approximation (LDA) of DFT, or some
of its semi-local variants (GGA, meta-GGA, etc.). This approach has been highly
successful, and routinely yields good results for global surface properties such
as work functions, surface energies, crystal-structure relaxation and reconstruction,
etc.~\cite{nekovee}

At a more basic level, however, some problems still remain to be solved,
concerning for instance the asymptotic behavior of the exchange-correlation ({\it xc}) 
potential of the widely used Kohn-Sham (KS) approach to DFT. In the LDA, this potential decays
exponentially when evaluated in the vacuum region, instead of the expected image-like
$\propto - \; e^2/z$ behavior.~\cite{almbladh} This qualitative failure of the LDA {\it xc} potential translates to a similar
failure of the position-dependent {\it xc} energy per particle,
$\varepsilon_{xc}({\bf r})$, which is defined through~\cite{gunnarsson}
\begin{equation}
E_{xc}[n] = \int n({\bf r}) \; \varepsilon_{xc} ({\bf r}) \; d{\bf r},
\label{definition}
\end{equation}    
with $E_{xc}[n]$ being the {\it xc}-energy contribution to the universal energy
functional of DFT, and $n({\bf r})$ representing the electron density. Three aspects of
Eq.~(\ref{definition}) are worth emphasizing: {\it (i)} it represents the basic expression
for the LDA, in which the exact $\varepsilon_{xc}({\bf r})$ of an arbitrary inhomogeneous
electron system is replaced at each point ${\bf r}$ by that of a homogeneous electron gas at
the local density $n({\bf r})$, and for a plethora of generalizations of the LDA,~\cite{tao} 
 {\it (ii)} since $E_{xc}[n]$ can be split as the sum of exchange ($E_x[n]$) and correlation
($E_c[n]$) contributions, one can write 
$\varepsilon_{xc}({\bf r}) = \varepsilon_{x}({\bf r}) + \varepsilon_{c}({\bf r})$, and {\it (iii)} the 
position-dependent {\it xc} energy per particle $\varepsilon_{xc}({\bf r})$ 
entering Eq.~(1) is not unique. One can always add to $\varepsilon_{xc}({\bf r})$ an arbitrary 
function $\varepsilon({\bf r})$ with the condition that weighted by the electron 
density $n({\bf r})$ integrates to zero. Here we have chosen $\varepsilon_{xc}({\bf r})$ to 
represent the interaction between a given electron at ${\bf r}$ and its {\it xc} hole.

The goal of this work is to provide exact analytical and numerical calculations of the exact-exchange 
$\varepsilon_x({\bf r})$ for jellium slabs and the semi-infinite (SI) jellium.
In particular, we analyze the asymptotic behavior of the exact
 $\varepsilon_{x}({\bf r})$ in the vacuum region far away from the surface, and we find that
there is a qualitative difference between $\varepsilon_x^{\text{Slab}} (z \rightarrow \infty)$ and
$\varepsilon_x^{\text{SI}} (z \rightarrow \infty)$: both exhibit an image-like behavior
of the form $-a\,e^2/z\,(a > 0)$, but with a coefficient $a$ that while in the case of jellium slabs is
universal and equal to 1/2 in the case of a semi-infinite jellium depends on the density of the bulk
material and only approaches 1/2 in the low-density limit. The results reported here should help to
settle the still controversial issue of the asymptotic behavior of the position-dependent {\it xc} energy
per particle and KS {\it xc} potential at metal surfaces.~\cite{gunnarsson,ss,ssb,nastos}

Besides, being the results presented here exact at the exchange level, they should also serve as a benchmark
against to which DFT {\it xc} calculations could be confronted, and hopefully improved,
once reduced to their exchange-only version. In this context, very recently 
Luo {\it et al.}~\cite{luo} have used a HF scheme to report self-consistent calculations of the surface energy
and work function of jellium slabs.

The rest of the paper is organized as follows: In Section II we present the
general theoretical background for both jellium slabs and the semi-infinite jellium,
and we derive exact analytical expressions for the position-dependent exchange energy per particle in the vacuum region far away from the surface. Numerical calculations that are valid at all positions, from the bulk region to the vacuum, are reported in Section III. Section IV is devoted to the conclusions.

\section{Jellium slabs and the semi-infinite jellium: the exact asymptotic behavior}

\subsection{Jellium slabs}

In the case of jellium slabs, with the discrete character of the positive ions inside the slab being replaced by a
uniform distribution of positive charge (the jellium background), the positive jellium density is  
\begin{equation}
n_{+}^{\text{Slab}}(z)=\overline{n}\,\theta (-z) \theta (d+z),
\label{jellium}
\end{equation}
which describes a slab of width $d$, number density $\overline{n}$,~\cite{note0}
and jellium edges at $z=-d$ and $z=0$. $\theta(z)$ represents the Heaviside
step function: $\theta (z)=1$ if $z > 0$ and $\theta (z)=0$ if $z < 0$. 
The size of the slab is infinite in the $x-y$ plane. The jellium slab is taken to be 
invariant under translations in the
$x-y$ plane, so the KS eigenfunctions can be rigorously factorized as follows 
\begin{equation}
\varphi_{i,{\bf k}}({\bf r})=\frac{e^{i{\bf k\cdot} \bm{ \rho }}}{\sqrt{A}}\,
\xi_{i}(z),
\label{KSfunctions}
\end{equation}
where $\bm{\rho}$ and ${\bf k}$ are the in-plane coordinate and
wave-vector, respectively, and $A$ represents a normalization area in the $x-y$ plane.
$\xi _{i}(z)$ are normalized spin-degenerate eigenfunctions for electrons in slab discrete levels (SDL)
$i$ $(i=1,2,...)$ with energies $\varepsilon_{i};$ they are the solutions of the effective one-dimensional KS equation

\begin{equation}
\widehat{h}_{\text{KS}}^{i}(z)\xi _{i}(z)=\left[ -\frac{\hbar ^{2}}{2m_{e}}
\frac{\partial ^{2}}{\partial z^{2}}+V_{\text{KS}}\left( z\right)
-\varepsilon _{i}\right] \xi _{i}(z)=0,
\label{KSequations}
\end{equation}
with $m_e$ being the electron mass. It is important to remark here that the factorization
of the 3D wave-function as proposed in Eq.~(\ref{KSfunctions}) is only valid for the
case of a {\it local} potential, as is the case of the KS implementation of DFT. On the other
hand, in the HF approximation the non-locality of the Fock potential introduces a coupling
between ${\bf k}$ and $i$ quantum numbers~\cite{luo}. As a consequence, HF numerical calculations are
more time-consuming than the ones presented here, either LDA, KLI~\cite{KLI}, or OEP.~\cite{grabo}

The local KS potential $V_{\text{KS}}(z)$ entering Eq.~(\ref{KSequations}) is the sum of
two distinct contributions: 
\begin{equation}
V_{\text{KS}}(z)=V_{\text{H}}(z)+V_{xc}(z),  \label{KSpotential}
\end{equation}
where $V_{\text{H}}(z)$ is the classical (electrostatic) Hartree potential,
given by~\cite{hartree.a} 
\begin{equation}
V_{\text{H}}(z)=-2\pi e^{2}\int_{-\infty }^{\infty }dz^{\prime }\left|
z-z^{\prime }\right| \left[ n^{\text{{Slab}}}(z^{\prime })-n_{+}(z^{\prime })\right].
\label{hartree}
\end{equation}
Here, $n^{\text{Slab}}(z)$ is the electron number density~\cite{dens.aa} 
\begin{equation}
n^{\text{Slab}}(z)=\frac{1}{2\pi }\sum_{i}^{occ.}\left( k_{F}^{i}\right) ^{2}\left| \xi
_{i}(z)\right| ^{2},  \label{density}
\end{equation}
where
$k_{F}^{i}=\sqrt{2m_{e}(\varepsilon_F -\varepsilon _{i})}/\hbar$,
and $\varepsilon_F = \varepsilon_F (\bar n,d) $ is the Fermi energy or chemical potential, 
which in turn is determined from the neutrality
condition for the whole system by the condition
$\sum_{i}^{occ.}(k_{F}^{i})^{2}=2\pi \,d\,\overline{n}$. $V_{xc}(z)$ is the
nonclassical {\it xc} potential, which is obtained as the functional derivative of
the {\it xc}-energy functional $E_{xc}[n(z)]$:~\cite{factor.a}
\begin{equation}
V_{xc}(z)\equiv \frac{1}{A}\frac{\delta E_{xc}[n(z)]}{\delta n(z)} \; .
\label{xcpotential}
\end{equation}

Both for the slab and semi-infinite [$d \rightarrow \infty$ limit of Eq.~(\ref{jellium}))]
geometries, and as a consequence of the translational symmetry
in the $x-y$ plane, Eq.~(\ref{definition}) simplifies to
\begin{equation}
E_{xc}[n] = A \int\limits_{-\infty }^{\infty } dz\,n(z)\,\varepsilon_{xc}(z),
\label{xcenergy}
\end{equation}
where $\varepsilon_{xc}(z)$ is the position-dependent {\it xc} energy per particle at plane $z$.
In the case of jellium slabs, the exchange-only contribution to $E_{xc}[n]$ (which is originated
in the Pauli exchange hole with all other correlation effects excluded) is known to be given by
the following expression:~\cite{reboredo} 
\begin{equation}
E_{x}^{\text{Slab}}[n]=-2e^2 A \sum_{i,j}^{occ.} k_{F}^{i} k_{F}^{j}
\int\limits_{-\infty }^{\infty }dz\int\limits_{-\infty }^{\infty
}dz^{\prime}\varphi_{i}(z,z^{\prime })\varphi _{j}
(z^{\prime },z)F_{ij}(z,z^{\prime }),
\label{xenergy}
\end{equation}
where $\varphi _{i}(z,z^{\prime })=\xi _{i}(z)^{*}\xi
_{i}(z^{\prime })$ and 
\begin{equation}
F_{ij}(z,z^{\prime })=\frac{1}{4\pi }\int\limits_{0}^{\infty }\frac{%
d\rho }{\rho }\frac{J_{1}(\rho k_{F}^{i})J_{1}(\rho
k_{F}^{j})}{\sqrt{\rho ^{2}+(z-z^{\prime })^{2}}} \; ,
\label{2}
\end{equation}
with $J_1(x)$ being the cylindrical Bessel function of first order.~\cite{abra}

Comparison of Eq.~(\ref{xcenergy}) and (\ref{xenergy}) yields the following expression for the
exchange-only contribution to $\varepsilon_{xc}^{\text{Slab}}(z)$: 
\begin{equation}
\varepsilon _{x}^{\text{Slab}}(z)=-\frac{2e^2}{n^{\text{Slab}}(z)}\sum_{i,j}^{occ.}k_{F}^{i}k_{F}^{j}\int
\limits_{-\infty}^{\infty}dz^{\prime }\varphi _{i}(z,z^{\prime })\varphi _{j}(z^{\prime
},z)F_{ij}(z,z^{\prime }),
\label{5}
\end{equation}
which can be interpreted as the energy due to the interaction of an electron at $z$ and its exchange-only
Pauli hole. In order to demonstrate that the exact-exchange energy per particle $\varepsilon_x^{slab}(z)$ 
of Eq.~(\ref{5})
 represents indeed the interaction between an electron at $z$ and its exact-exchange hole, we appeal to the following expression for the exchange-hole for our slab geometry\cite{rrp}
\begin{equation}
h_x({\bf r};{\bf r+R})=\frac{-1}{2(\pi \rho)^2 n^{\text{Slab}}(z)} \sum_{i,j}^{occ.}k^{i}_{F}k^{j}_{F}J_1(\rho k^{i}_{F})J_1(\rho k^{j}_{F})
\xi_{i}(z+Z)^{*}\xi_{i}(z)\xi_{j}(z+Z)\xi_{j}(z)^{*} ; ,
\label{xhole}
\end{equation}
which represents the density of the exchange hole at point ${\bf r+R}$ (observational point) due to the presence 
of an electron located at {\bf r}. Owing to the translational symmetry in the $x-y$ plane,
without loss of generality we have choose ${\bf r}=(0,z)$, ${\bf r+R}$ = ($\bm {\rho}$, $z+Z$).
Using Eq.~(\ref{xhole}), and defining $z^{\prime} = z+Z$, Eq.~(\ref{5}) may be rewritten as
\begin{equation}
\varepsilon_{x}^{\text{Slab}}(z) = \frac {e^2}{2} \int d \bm {\rho} \int dz^{\prime} \frac {h_{x}(z;\rho,z^{\prime})}
{\sqrt{\rho^2+(z-z^{\prime})^2}} \; ,
\label{physical}
\end{equation}
which justifies the physical interpretation of $\varepsilon_{x}^{\text{Slab}}(z)$ as the interaction energy of an electron located at $z$ and the ``charge distribution'' given by $h_{x}(z;\rho,z^{\prime})$.
Equations~(\ref{xhole}) and (\ref{physical}) can also be used as a sort of alternative definition 
of the $\varepsilon_{x}^{\text{Slab}}(z)$ investigated
in this work, as they solve the non-uniqueness of $\varepsilon_{x}^{\text{Slab}}(z)$ which results from
its definition through the exchange-only version of Eq.~(\ref{xcenergy}).\cite{tao01,tsp}  
 
\subsubsection{Single occupied slab discrete level}

In order to obtain the asymptotic behavior of $\varepsilon _{x}^{\text{Slab}}(z)$ at
$z\rightarrow \infty$, we first restrict our analysis to the case where there is one single occupied
SDL.~\cite{oneSDL} In this special case, in which $i=j=1$, Eq.~(\ref{5}) yields
[see also Eqs.~(\ref{density}) and (\ref{2})]:
\begin{equation}
\varepsilon _{x,1}^{\text{Slab}}(z)=- \; e^2 \; \int \limits_{-\infty}^{\infty} dz^{\prime }\left| \xi _{1}(z^{\prime
})\right| ^{2}\int\limits_{0}^{\infty }\frac{d\rho }{\rho }\frac{\left[
J_{1}(\rho k_{F}^{1})\right] ^{2}}{\sqrt{\rho ^{2}+(z-z^{\prime })^{2}}},
\label{7}
\end{equation}
or, equivalently (see Appendix):
\begin{equation}
\varepsilon _{x,1}^{\text{Slab}}(z)=-\frac{e^2}{2}\int\limits_{-\infty }^{\infty
}dz^{\prime }\frac{\left| \xi _{1}(z^{\prime })\right| ^{2}}{\left|
z-z^{\prime }\right| }\left[ 1-\frac{I_{1}(2k_{F}^{1}\left| z-z^{\prime
}\right| )}{k_{F}^{1}\left| z-z^{\prime }\right| }+\frac{L_{1}(2k_{F}^{1}%
\left| z-z^{\prime }\right| )}{k_{F}^{1}\left| z-z^{\prime }\right| }\right],
\label{8}
\end{equation}
with $I_{1}$ and $L_{1}$ being the modified Bessel and Struve functions,
respectively.~\cite{abra}

We note that Eq.~(\ref{8}) is valid for {\it all} $z,$ both inside and outside the jellium slab.
Also, the cancellation of $n^{\text{Slab}}(z)$ which occurs in passing from Eq.~(\ref{5}) to Eq.~(\ref{7})
allows for the numerical calculation of $\varepsilon _{x,1}^{\text{Slab}}(z)$ for arbitrarily large values
of $z$. 

\paragraph{Asymptotic behavior.}
For the slab geometry,
it is permissible (and rigorous) to take the asymptotic limit 
$k_{F}^{1}\left| z-z^{\prime }\right| \simeq z\,k_{F}^{1}\gg 1,$ although the
integral over $z^{\prime }$ runs from $-\infty $ to $+\infty .$ 
This is due to the fact that for a given $z$ the main contribution to
the integral in Eq.~(\ref{8}) comes from values of $z^{\prime }$ inside the slab
($-d\lesssim z^{\prime }\lesssim 0$), as $\xi_{1}(z)$ decays exponentially
one or two $\lambda _{F}$'s from each jellium edge. 
 
At this point, we define $F(x)\equiv 1-I_{1}(2x)/x+L_{1}(2x)/x$ and use the
asymptotic expansions of $I_{1}$ and $L_{1}$ in the limit $x\gg 1.$ We
obtain~\cite{abra}
\begin{equation}
F(x\gg 1)\rightarrow 1-\frac{2}{\pi x}+\frac{1}{2x^{3}}-... \; \; .
\label{9}
\end{equation}
Hence, as $z\,k_{F}^{1}\gg 1$, Eq.~(\ref{8}) yields the following
asymptotic behavior: 
\begin{equation}
\varepsilon _{x,1}^{\text{Slab}}(z\rightarrow \infty )\rightarrow -\frac{e^{2}}{2z}\left( 1+\frac{
\beta_{1} }{z}+\frac{\gamma_{1} }{z^{2}}+...\right) ,  \label{sdpasymp}
\end{equation}
where $\beta_{1} (d,r_{s}) =\overline{z}^1 (d,r_{s}) - 2/\left[ \pi k_{F}^{1}(d,r_{s})\right]$ and
$\gamma_{1}(d,r_{s})=  \overline{z^{2}}^1(d,r_{s}) -
4\, \overline{z}^1/\left[ \pi k_{F}^{1} (d,r_{s})\right]$,
mean values being defined here as
$\overline{O}^{i}=\int \xi _{i}(z)^{*}O(z)\,\xi _{i}(z)\,dz.$ The $r_{s}$ dependence of $\overline{z}^1(d,r_{s})$
and $\overline{z^{2}}^1(d,r_{s})$ comes from the self-consistent KS wave-functions $\xi_{i}(z)$,
which for a given $d$ are different for different values of the slab density dictated by $r_{s}$. For details
on the derivation of Eq.~(\ref{sdpasymp}) from Eq.~(\ref{8}), we refer to the Appendix. 

\subsubsection{General situation}

For the general situation where more than one SDL is occupied, we obtain the asymptotic limit of
Eq.~(\ref{5}) by using the fact that for $z \rightarrow \infty$ {\it (i)}
the electron density is dominated by the slowest decaying KS orbital, which corresponds to the
{\it highest occupied} SDL ($i=m$), and {\it (ii)} the numerator of Eq.~(\ref{5}) is dominated by the term $i=j=m$,
since all $\xi_{i}(z)$ with $i\neq m$ decay exponentially two or three $\lambda _{F}$'s from 
each jellium edge. Hence, in the vacuum region far away from the surface we find:
\begin{equation}
n^{\text{Slab}}(z\rightarrow \infty )\rightarrow \frac{(k_{F}^{m})^2}{2\pi} 
 \left| \xi _{m}(z)\right| ^{2}
\label{13}
\end{equation}
and
\begin{equation}
\varepsilon_{x}^{\text{Slab}}(z\rightarrow \infty ) \rightarrow -\frac{4 \pi e^2}{ 
\left| \xi _{m}(z)\right| ^{2}}  \int\limits_{-\infty}^{\infty}
dz^{\prime }\varphi _{m}(z,z^{\prime })
\varphi _{m}(z^{\prime},z)F_{mm}(z,z^{\prime }),
\label{14}
\end{equation}
or, equivalently [see Eq.~(\ref{2})]:
\begin{equation}
\varepsilon _{x}^{\text{Slab}}(z\rightarrow \infty ) \rightarrow - \; e^2 \int\limits_{-\infty}^{\infty} dz^{\prime }\left| \xi _{m}(z^{\prime
})\right| ^{2}\int\limits_{0}^{\infty }\frac{d\rho }{\rho }\frac{\left[
J_{1}(\rho k_{F}^{m})\right] ^{2}}{\sqrt{\rho ^{2}+(z-z^{\prime })^{2}}}\;.
\label{15}
\end{equation}

Finally, following the same procedure as in the case of a single occupied SDL, we find:
\begin{equation}
\varepsilon_{x}^{\text{Slab}}(z\rightarrow \infty )\rightarrow -\frac{e^{2}}{2z}\left( 1+\frac{
\beta_{m}}{z}+\frac{\gamma_{m}}{z^{2}}+...\right) ,  \label{sdpasymp2}
\end{equation}
where $\beta_{m}(d,r_{s}) =\overline{z}^m(d,r_{s}) - 2/\left[ \pi k_{F}^{m}(d,r_{s})\right]$ and
$\gamma_{m} (d,r_{s}) =  \overline{z^{2}}^m(d,r_{s}) -
4\, \overline{z}^m(d,r_{s})/\left[ \pi k_{F}^{m}(d,r_{s})\right]$.~\cite{note.bb}

This result represents a straightforward generalization of the result presented above
[Eq.~(\ref{sdpasymp})] for the case of one single occupied SDL. At this point,
it is interesting to note that the leading contribution to 
$\varepsilon_x^{\text{Slab}}(z \rightarrow \infty) \rightarrow - \; e^2/2z$ can be 
easily obtained directly from Eqs.~(\ref{7}) or (\ref{15}),
by approximating the argument inside the square root by $z$
(in the large $z$ limit), and using the normalization of the KS orbitals $\xi_i(z)$
and the identity $\int \limits_{0}^{\infty} dx J_1(x)^2 / x = 1/2$. 

By considering a slab of thickness sufficiently large to make the energy spectrum {\it continuous},
Solamatin and Sahni~\cite{ssb} reached the conclusion that far away from the slab the so-called Slater potential $V_S(z)$ [which is twice the exchange energy per particle: $V_S(z)=2\varepsilon_x(z)$] decays as $-e^2/z^2$, in contrast with
the asymptotic structure dictated by Eq.~(\ref{sdpasymp2}). This result is, however, not correct due to the fact that for a finite jellium slab (no matter how thick it is) the slab {\it intrinsic} discrete spectrum [corresponding to the eigenvalues $\varepsilon_i$ entering Eq.~(\ref{KSequations})] can never be replaced by a continuous one.

Equation~(\ref{sdpasymp2}) leads us to the conclusion that in the vacuum region of a finite jellium slab and at distances from the surface that are large compared to $1/k_F^m$ (which is typically larger than the slab thickness $d$),
$\varepsilon_{x}^{\text{Slab}}(z \rightarrow \infty) \rightarrow - \; e^{2}/2z$, which is exactly half the corresponding KS exact-exchange potential $V_x(z\to\infty)\rightarrow - \; e^{2}/z$.\cite{hpr} Hence, as in the case of finite systems,\cite{gllb} the Slater potential $V_S(z)$ of jellium slabs [or, equivalently, twice the exchange-energy per particle
$\varepsilon_x(z)$] embodies the asymptotics of the KS exchange potential $V_x(z)$.

In contrast, Solamatin and Sahni~\cite{ss,ssb} concluded that in the case of a semi-infinite jellium  only {\it half} the Slater potential embodies the asymptotics of the KS exchange potential, i.e., 
$V_x(z\to\infty)=\varepsilon_x(z\to\infty)$; but Nastos~\cite{nastos} claimed that $V_x(z\to\infty)=2\varepsilon_x(z\to\infty)$, so there is still something remaining to be clarified on this issue. Work along these lines is now in progress.\cite{notesahni0} 

\subsection{Two-dimensional electron gas}

The exchange energy of a strict two-dimensional (2D) electron gas can be obtained from that of a jellium slab 
with a single occupied SDL [Eq.~(\ref{xenergy}) with $i=j=1$], by first performing the one-dimensional non-uniform scaling~\cite{pollak}
\begin{equation}
n_{\lambda}^{\text{Slab}}(z) = \lambda \; n^{\text{Slab}}(\lambda z) \; ,
\end{equation}
and then taking the limit as $\lambda\to\infty$. The scaling above preserves the total number of electrons.
Noting that for a single occupied SDL the jellium-slab exchange energy takes the following form
\begin{equation}
E_{x}^{\text{Slab}}[n]=- \frac{2 \pi e^2 A}{\left( k_{F}^{1} \right)^2}
\int\limits_{0}^{\infty }\frac{d \rho}{\rho} \left[ J_{1}(k_{F}^{1}\rho)\right]^2 \int\limits_{-\infty }^{\infty} 
dz \int\limits_{-\infty }^{\infty} dz^{\prime} 
\frac {n^{\text{Slab}}(z) \; n^{\text{Slab}}(z^{\prime})} {\sqrt{\rho^2+(z-z^{\prime})^2}},
\label{explicit}
\end{equation}
where
\begin{equation}
n^{\text{Slab}}(z)=\frac {\left( k_{F}^{1}\right)^{2}}{2\pi} \left| \xi
_{1}(z)\right|^{2},
\label{2Ddensity}
\end{equation}
we find:
\begin{equation}
E_x^{\text{2D}} \equiv \lim_{\lambda\to\infty}E_x^{\text{Slab}}[n_{\lambda}^{\text{Slab}}]
=-N\,\frac {4}{3 \pi} e^2 \; k_F^1,
\label{2Dlimit}
\end{equation}
where $N = A \left( k_F^1 \right)^2 / (2 \pi)$ represents the total number of electrons. 
Previously, this scaling limit had been formulated in a different way, resulting in the 
much generous constraint that the exchange energy per particle in the 2D ($\lambda \rightarrow \infty$) limit should be greater than
$- \; \infty$.~\cite{pollak,cpp,constantin} It is interesting to note that the exchange energy
functional as given by Eq.~(\ref{explicit}) is an {\it explicit} functional of the density, which
is only possible in this single occupied SDL case, due to the simple (invertible) relation between density
and wave-function, as given by Eq.~(\ref{2Ddensity}). In the general, many SDL occupied case,
Eq.~(\ref{2Ddensity}) is replaced by Eq.~(\ref{density}), the direct inversion from wave-functions
to density is not feasible anymore, and the exchange energy functional is an explicit
functional of the KS orbitals, but an {\it implicit} functional of the density, as in Eq.~(\ref{xenergy}). 

We note at this point that the exchange energy of a strict 2D electron gas can also be obtained directly from 
Eq.~(\ref{7}) through the replacement
$\xi_{1}(z')\rightarrow \sqrt{\delta (z')},$ with $\delta (z')$ being the Dirac
delta function, and taking $z=0$: 
\begin{equation}
E_x^{\text{2D}}=-N\,e^2\int\limits_{0}^{\infty }\frac{d\rho }
{\rho^2}\,\left[ J_{1}(\rho k_{F}^{1})\right]^{2}
=-N\,\frac{4}{3\pi }e^2 k_{F}^{1}.
\label{11}
\end{equation}
Either from Eq.~(\ref{2Dlimit}) or (\ref{11}), we find for the exchange energy per particle
of the strict 2D homogeneous electron gas the well-known result
$\varepsilon_x^{\text{2D}} = E_x^{\text{2D}}/N = -(4/3 \pi) e^2 k_F^{1} \; $. \cite{GV}
  
\subsection{Semi-infinite jellium}
  
In the case of a semi-infinite jellium, a half-space filled with a uniform distribution of positive charge (the jellium background), the jellium density is
\begin{equation}
n_{+}^{\text{SI}}(z)=\overline{n}\, \theta \left(- z \right),
\label{jellium.semi}
\end{equation}
with the jellium edge at $z=0$ defining the surface of a metal. 
As in the case of jellium slabs, the semi-infinite jellium is invariant under translations in the $x-y$ plane, so the KS eigenfunctions can be factorized as follows 
\begin{equation}
\varphi_{k_z,{\bf k}}({\bf r})=\frac{e^{i{\bf k\cdot}\bm{\rho }}}{\sqrt{A}}\,
\frac {\xi_{k_z}(z)}{\sqrt{L}},
\label{KSfunctions.semi}
\end{equation}
where $\bm{\rho}$ and ${\bf k}$ are the in-plane coordinate and
wave-vector, respectively, and $A$ ($L$) represents a normalization area (length).
$\xi _{k_z}(z)$ are spin-degenerate eigenfunctions for electrons with a continuous energy spectrum 
$\varepsilon_{k_z} = V_{\text{KS}}(- \infty) + (\hbar k_{z})^{2}/2m_e$ ($k_z$ is a continuum quantum number). They are the solutions of the effective one-dimensional
KS equation 
\begin{equation}
\widehat{h}_{\text{KS}}^{k_z}(z)\xi _{k_z}(z)=\left[ -\frac{\hbar ^{2}}{2m_{e}}
\frac{\partial ^{2}}{\partial z^{2}}+V_{\text{KS}}\left( z\right)
-\varepsilon _{k_z}\right] \xi _{k_z}(z)=0.
\label{KSequations.semi}
\end{equation}
The KS potential $V_{\text{KS}}(z)$ is given by Eq.~(\ref{KSpotential}), as in the case of a jellium slab but with the slab electron density of 
Eq.~(\ref{density}) being replaced by the SI electron density
\begin{equation}
n^{\text{SI}}(z)=\frac{1}{4 \pi^{2}}\int\limits_{-k_{F} }^{k_{F}} (k_{F}^{2}-k_{z}^{2})
\left| \xi_{k_{z}}(z)\right|^{2} dk_z.
\label{densidad.semi}\\
\end{equation}

In the case of a semi-infinite jellium, the position-dependent exchange energy per particle at plane $z$ is given by the following expression: 
\begin{equation}
\varepsilon_{x}^{\text{SI}}(z)=-\frac{e^2}{2 \pi^2 n^{\text{SI}}(z)}
\int\limits_{-k_{F}}^{k_{F}} dk_z \int\limits_{-k_{F} }^{k_{F}} dk_{z}^{'} (k_{F}^{2}-k_{z}^{2})^{1/2}
(k_{F}^{2}-k_{z}^{'2})^{1/2} \int\limits_{-\infty}^{\infty}
dz^{\prime }\varphi _{k_z}(z,z^{\prime })\varphi _{k_{z}^{'}}(z^{\prime
},z)F_{k_z k_{z}^{'}}(z,z^{\prime }),
\label{5.semi}
\end{equation}
where $\varphi_{k_z}(z,z^{\prime })=\xi _{k_z}(z)^{*}\xi_{k_z}(z^{\prime })$,
and $F_{k_z k_z^{'}}(z,z^{\prime })$ is of the form of Eq.~(\ref{2}) but with $k_F^i$ ($k_F^j$) being replaced by $[k_F^2-k_z^2]^{1/2}$ ($[k_F^2-k_z^{'2}]^{1/2}$).

\paragraph{Asymptotic behavior.} 
The derivation of the asymptotic limit of $\varepsilon_x^{\text{SI}}(z)$ is more
delicate than in the case of jellium slabs, due to the fact that in the present case we have a continuous energy spectra. Hence, the crucial argument that we have used to derive the asymptotic behavior for jellium slabs, concerning the fact that in that case the highest occupied SDL dominates in the vacuum region far away from the surface, is not so transparent when the spectrum is continuous.
Stated in other words, contributions to the asymptotic position-dependent exchange energy of 
Eq.~(\ref{5.semi}) come indeed from values of $k_z$ and $k_z^{'}$ that approach $k_F$, but not necessarily {\it only} from the highest occupied value, i.e., from
$k_z = k_z^{'} = k_F$.

The asymptotic behavior of Eq.~(\ref{5.semi}) was analyzed by Nastos~\cite{nastos}
by assuming that at $z\to\infty$ the KS potential $V_{\text{KS}}(z)$ takes the 
image-like form $V_{\text{KS}}(z \rightarrow \infty) \rightarrow - \; \alpha_{\text{KS}}\,e^2/ z$, 
with $\alpha_{\text{KS}}$ positive, but otherwise arbitrary. One finds that in the vacuum region far away from the surface
($z\rightarrow\infty$) the KS orbitals $\xi_{k_{z}}$ can be expanded with respect to the KS orbital at $k_z=k_F$ as follows\cite{nastos,qs}
\begin{equation}
\xi _{k}(z \rightarrow \infty) \rightarrow \xi _{k_{F}}(z \rightarrow \infty) 
e^{-\alpha z (k_{F}-k) } \; ,
\label{aprox.nastos}
\end{equation}
with 
\begin{equation}
\xi _{k_F}(z \rightarrow \infty ) \propto e^{-z \sqrt{2m_{e}W / \hbar^2}} (2z \sqrt{2W})^{\alpha_{\text{KS}}/ \sqrt{2m_ea_0^2W/\hbar^2}},
\label{aprox.nastos1}
\end{equation}
$\alpha$ standing for the square root of the ratio between the Fermi energy and the work function $W$ ($\alpha ^{2} = \varepsilon _{F} / W$). By introducing 
Eqs.~(\ref{aprox.nastos}) and (\ref{aprox.nastos1}) into Eqs.~(\ref{densidad.semi}) and
(\ref{5.semi}), one finds\cite{nastos,qs}
\begin{equation}
n^{\text{SI}} (z\rightarrow \infty) \rightarrow \frac{3 \overline{n}}{4(\alpha k_F z)^2} 
\left| \xi_{k_F}(z\rightarrow \infty ) \right|^2 \; ,
\label{density-semi}
\end{equation}
and
\begin{equation}
\varepsilon _{x}^{\text{SI}}(z\rightarrow \infty )
\rightarrow -\frac{\pi + 2 \alpha \ln (\alpha)}{2 \pi (1+\alpha^2)} \frac{e^2}{z}.
\label{semi}
\end{equation}
Furthermore, the asymptote of Eq.~(\ref{semi}) does not depend on the actual form of $\xi_{k_F} (z \rightarrow \infty)$. This is due to a cancellation, when $z$ is large,
of the orbitals $\xi _{k_F}(z \rightarrow \infty )$ entering the numerator and denominator of 
Eq.~(\ref{5.semi}). This is the reason why the leading term in the expansion of $\varepsilon_x^{\text{SI}}(z \rightarrow \infty)$
is independent of $\alpha_{\text{KS}}$. That means that the result remains valid also in the absence of
this image-like contribution, i.e., even assuming that the KS potential $V_{\text{KS}}(z)$ entering
Eq.~(\ref{KSequations}) decays exponentially as $z \rightarrow \infty$.

The asymptotic behavior dictated by Eq.~(\ref{semi}) was obtained independently by Solamatin and Sahni 
using a somehow less general, but otherwise quite different approach.~\cite{ss,ssb,notesahni} 
They approximated the KS potential $V_{\text{KS}}(z)$ entering
Eq.~(\ref{KSequations.semi}) by a finite-linear-potential model at the interface region, 
and used the corresponding orbitals in each of the three regions where the model was defined: 
trigonometric functions (in the bulk region), Airy functions (near the surface), and exponential decaying functions (in the vacuum). 
As the only thing that matters in obtaining the asymptote of Eq.~(\ref{semi}) is 
the correct expansion of the KS orbitals with respect to the KS orbital at $k_z=k_F$ [as given by 
Eq.~(\ref{aprox.nastos})], and with this general expansion 
being fulfilled also within the finite-linear-model potential used in
Refs.~\onlinecite{ss} and \onlinecite{ssb}, Solamatin and Sahni obtained Eq.~(\ref{semi}) which is valid in general.
It is also worth of address the fact that the result of Eq.~(\ref{semi}) is in contrast with the asymptotic
behavior $\varepsilon_x(z \rightarrow \infty) \rightarrow - \; e^2/4z$ that one obtains for the
exchange energy per particle in the case of the Airy edge electron gas.~\cite{KM} This is
due to the fact that the asymptotic behavior of the solutions of the Airy edge gas is different from the
one given by Eq.~(\ref{aprox.nastos}), as for this model the potential increases linearly with
distance in the vacuum region, instead of approaching a constant value. An analysis of the so-called
Pauli and lowest-order correlation-kinetic components of the exchange energy per particle
$\epsilon_x(z\to\infty)$ can be found in Ref.~\onlinecite{qs}.
 
\section{Numerical results}

All the numerical calculations presented below have been carried out by ignoring 
all correlation effects (beyond the Pauli exchange). By this we mean that
the {\it xc} potential $V_{xc}(z)$ entering Eq.~(\ref{KSequations}) has been 
replaced by the exchange-only contribution $V_{x}(z)$, disregarding $V_{c}(z)$, 
both for jellium slabs and for the semi-infinite jellium. In the case of jellium slabs, 
$V_x(z)$ and the corresponding KS orbitals of Eq.~(\ref{KSequations}) can be obtained 
through the solution of the discrete version of the $x$-only optimized effective 
potential (OEP) method, as given for example by Eqs.~(14) or (20) of Ref.~\onlinecite{hpp}. 
This code is feasible and we have at our disposal these self-consistent exact-exchange (OEP) 
KS orbitals and exchange potentials.~\cite{hpp,hpr} Alternatively, an approximate way to obtain $V_x(z)$ and the corresponding 
KS orbitals (the exchange-only LDA orbitals) is to replace the actual exchange potential 
$V_{x}(z)$ by the exchange potential of a uniform electron gas at the local electron 
density $n(z)$, i.e., $V_{x}^{\text{LDA}}(z)=-\left[6n(z)/\pi \right]^{1/3}$ [hartrees]. 

\begin{figure}
\includegraphics*[scale=0.5,angle=-90]{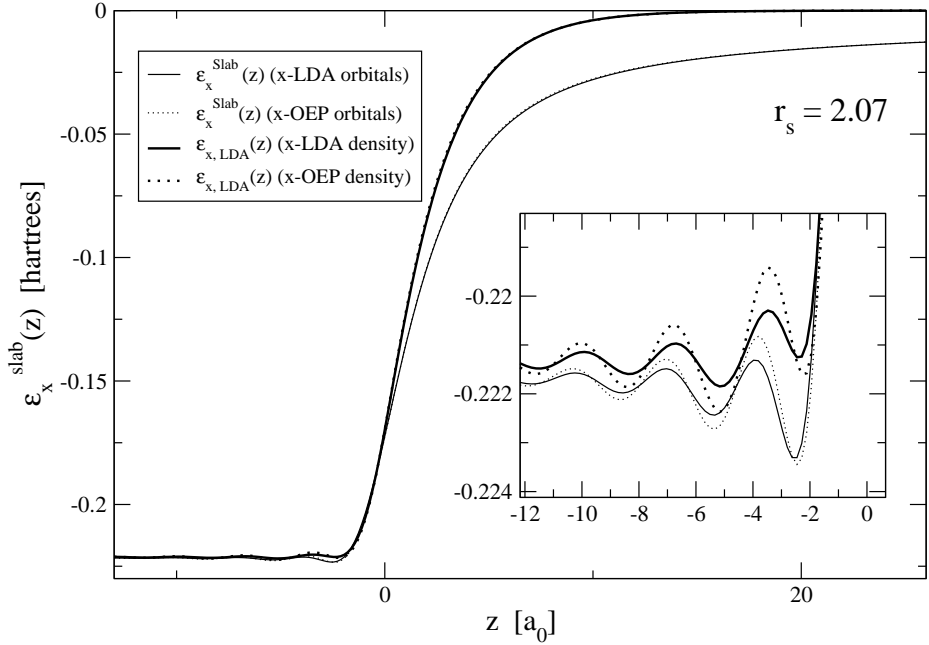}
\caption{$\varepsilon_{x}^{\text{Slab}}(z)$ for $r_s = 2.07$ and slab width
$d = 4.3 \; \lambda_F$. The curves denoted ``Slab'' have been evaluated from
Eq.~(\ref{5}), by using exchange-only LDA orbitals (solid line) and
exact-exchange (OEP) orbitals (dotted line). The curves denoted "LDA"  
have been evaluated from the well-known LDA formula 
$\varepsilon_{x,\text{LDA}}(z) = -(3/4 \pi) [3 \pi^2 n(z)]^{1/3}$ [hartrees],
with $n(z)$ being the exchange-only self-consistent electron density 
obtained with the use of exchange-only LDA orbitals (wide solid line) and
exact-exchange (OEP) orbitals (wide dotted line). Inset: enlarged view of the bulk region near the surface. The bulk value 
of $\varepsilon_x$ for this electron density is $- \; 0.22134 \; (e^2/a_0)$.}
\label{f1}
\end{figure}

Fig.~1 shows a comparison of (i) the well-known LDA exchange energy per particle 
$\varepsilon_{x,\text{LDA}}(z)= -(3 / 4 \pi) \left[ 3 \pi^2 n(z)\right]^{1/3}$ [hartrees],
with $n(z)$ being the self-consistent electron density 
obtained with the use of either exchange-only LDA orbitals (wide solid line) or
exact-exchange (OEP) orbitals (wide dotted line), with (ii) the exact-exchange energy per particle
$\varepsilon_x^{\text{Slab}}(z)$ of Eq.~(\ref{5}) obtained by using, as before, either exchange-only LDA orbitals
(solid line) or exact-exchange (OEP) orbitals (dotted line).
It is important to note that while both alternative evaluations of $\varepsilon_{x,\text{LDA}}(z)$  
fail badly in the vacuum region where the actual exchange energy per particle exhibits an image-like asymptotic 
behavior, the use of LDA orbitals in Eq.~(\ref{5}) results in an exchange energy per 
particle (solid line) that on the scale of the figure is nearly identical to the exact 
fully-self-consistent result (dotted line). Small differences introduced by the use of LDA orbitals 
(see the inset) are mostly localized in the bulk region near the surface, where 
Friedel-like oscillations appear to be too weak in this approximation.

The numerical methods that we have used to obtain exact-exchange (OEP) orbitals and exchange-only LDA orbitals suffer from instabilities in the vacuum region far from the surface. As in this region exchange-only LDA orbitals are stabler than their OEP counterparts and the results presented in this work do not depend significantly on whether exact-exchange (OEP) or exchange-only LDA orbitals are used in Eq.~(\ref{5}) (see Fig.~1), all the calculations presented below have been obtained with the use of exchange-only LDA orbitals.
We emphasize, however, that this is not a crucial approximation, and that the magnitude
of the error that it introduces is given by the almost indistinguishable difference
between the full and dotted lines in Fig.~1.
 
The numerical self-consistent calculations presented in Fig.~1 and in the 
remaining of this section (which are all obtained from either Eq.~(\ref{5}) 
or Eq.~(\ref{5.semi}) with the use of exchange-only LDA orbitals) have been performed 
as follows. For jellium slabs, two infinite barriers have been located in 
the vacuum region far enough away from the two surfaces, in such a way 
that all the numerical results be independent of their precise location,~\cite{note0a} 
and the KS equations have been solved through a straightforward discretization in 
real space along the one-dimensional coordinate $z$. In the case of the semi-infinite 
jellium, the KS equations have been solved by following the general procedure 
introduced by Lang and Kohn.~\cite{langkohn} This consists of defining three regions 
for the solution of the KS equation: far-left (bulk region), central (a few $\lambda_F$'s to 
the left and to the right of the jellium edge), and far-right (vacuum region). In the bulk 
region, the KS eigenfunctions are taken to be of the form $\xi _{k}(z) =\sin(kz-\gamma_{k})$, 
where $\gamma_{k}$ are phase shifts, and this fixes an overall normalization constant.
In the central region, we define a mesh of $N$ points between
$z_1$ and $z_N$ $(z_1 < z_N )$, the first point $z_1$ being chosen 
far enough from the jellium edge in the bulk so that the Friedel 
oscillations can be neglected, and the outer point $z_N$ being chosen to be far 
enough from the jellium edge into the vacuum so that the effective one-electron 
potential is negligibly small. 
Since $V_{\text{KS}}(z) \sim 0$ for $z \geq z_N$, the
orbitals can be approximated as $\xi_{k}(z_N) = a \, e^{-k^{*}z}$ where $a$ is a
constant and 
$k^{*} = (-2 m_e \varepsilon_{k_z} / \hbar^2)^{1/2}$. The KS orbitals at the 
mesh points are calculated by using the Numerov integration procedure.~\cite{liebsch}
As in the vacuum region the orbitals follow exponential form, it is numerically 
most stable to integrate them inwards, so the Numerov integration procedure in this
case is given by:
\begin{equation}
\xi_{k}(z_{i-1})= \frac{2+10 h (V_{\text{KS}}(z_{i})-\varepsilon_{k_z})}{1- h
(V_{\text{KS}}(z_{i-1})-\varepsilon_{k_z})} \xi _{k}(z_{i}) - 
 \frac{1- h (V_{KS}(z_{i+1})-\varepsilon_{k_z})}{1- h
(V_{\text{KS}}(z_{i-1})-\varepsilon_{k_z})} \xi _{k}(z_{i+1}),
\label{numerov}
\end{equation}
where $h=(z_{i+1}-z_{i})^{2} / 12$.
Finally, matching the KS orbitals in the central region 
with the corresponding analytical expression in the bulk region [$\xi_{k}(z)=\sin(kz-\gamma_{k})$]
determines the constant $a$.

\subsection{Jellium slabs}    

\begin{figure}
\includegraphics*[scale=0.5,angle=-90]{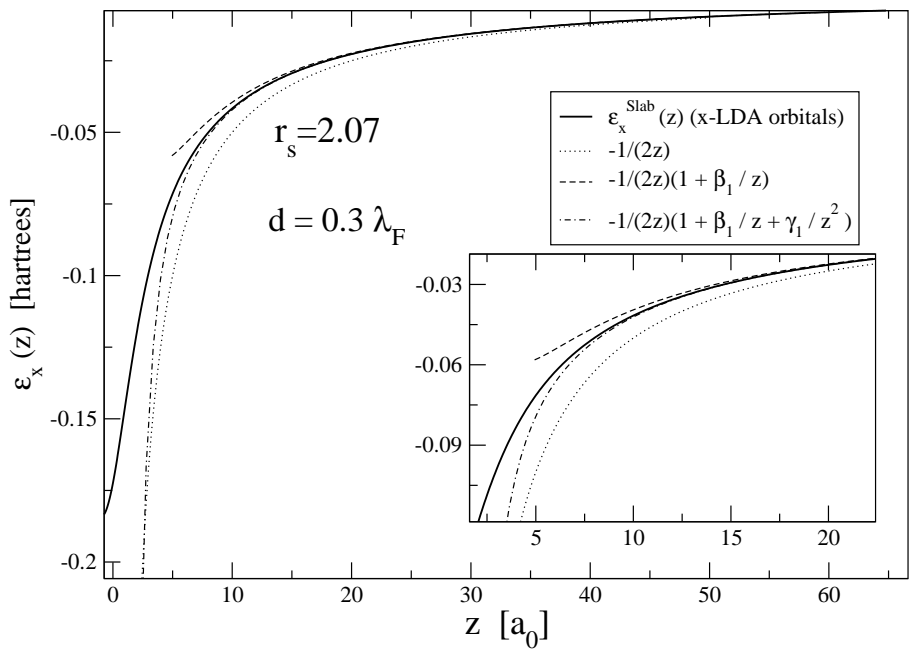}
\caption{Position-dependent exchange-energy per particle for a thin jellium slab with $r_s=2.07$
and $d = 0.3 \; \lambda_F \approx 2 a_0$. Solid line: full numerical calculation of Eq.~(\ref{5}). Dashed-dotted lines: asymptote of 
Eq.~(\ref{sdpasymp}); the dashed (dotted) lines represent the asymptote of
Eq.~(\ref{sdpasymp}) with the last term (last two terms) neglected. Inset: enlarged view of the asymptotic region.}
\label{f2}
\end{figure}

In Fig.~2, we consider a thin jellium slab with $r_s=2.07$ (corresponding to the average electron density of Al) and $d=0.3\,\lambda _{F}$. 
This slab contains {\it one} single occupied SDL, so that we compare our full numerical calculation of Eq.~(\ref{5}) (solid line) 
with the asymptote of 
Eq.~(\ref{sdpasymp}) (dashed-dotted line). We see that $\varepsilon_x^{\text{Slab}}(z)$ reaches 
Eq.~(\ref{sdpasymp}) at about one
$\lambda_F$ from the jellium edge and reaches the asymptote $-e^2/2z$ at a few Fermi wavelengths
($\sim 5-6\lambda_F$) from the surface.

\begin{figure}
\includegraphics*[scale=0.5,angle=-90]{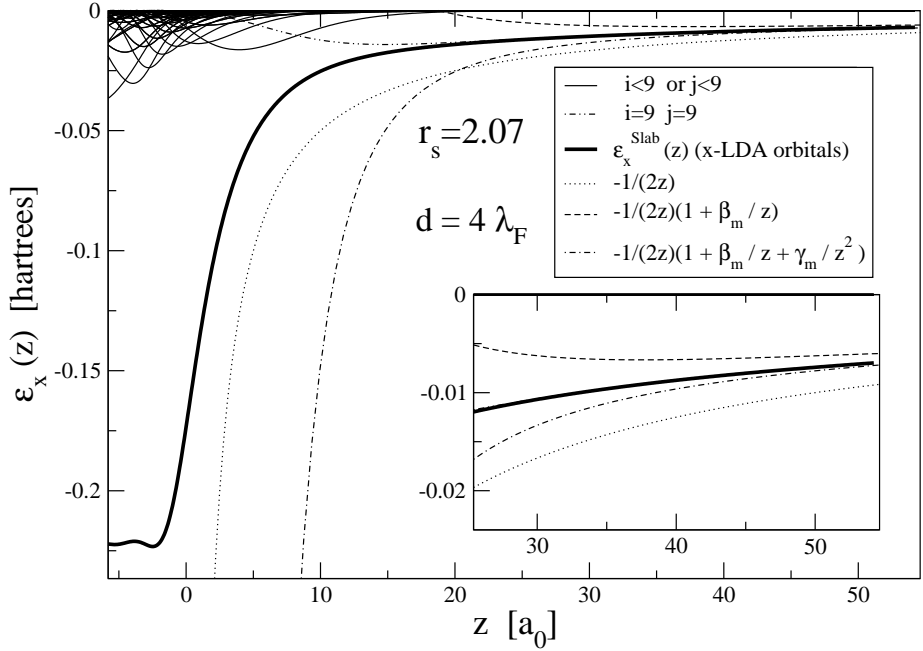}
\caption{Same as Fig.~2, but for a slab with $d = 4 \; \lambda_F$, with nine 
SDL occupied. Full thick line, $\varepsilon_{x}^{\text{Slab}}(z)$ from Eq.~(\ref{5});
dotted, dashed, and dashed-dotted lines, asymptotic expansions from Eq.~(\ref{sdpasymp2}).
The contribution from each pair $(i,j)$ of occupied SDL to the total ${\varepsilon_x}^{\text{Slab}}(z)$
are represented with thin full lines, except for the last contribution ($i = j = 9$). Inset:
enlarged view of the asymptotic region.}
\label{f3}
\end{figure}

In Fig.~3, we consider a jellium slab with $r_s=2.07$ and $d=4\,\lambda_{F}$. For this particular case,
{\it nine} SDL's are occupied, i.e., $\varepsilon_9<\varepsilon_F<\varepsilon_{10}$, so we compare our full numerical calculation of 
Eq.~(\ref{5}) (solid line) with the asymptote of Eq.~(\ref{sdpasymp2})
(dashed-dotted line). The main message of this figure is that {\it (i)} in the vacuum region 
far away from the surface $\varepsilon_x^{\text{Slab}}(z)$ is dominated by the 
term $i=j=m$ (dashed-dotted-dotted line), and {\it (ii)} $\varepsilon_x^{\text{Slab}}(z)$ reaches 
the asymptote $-e^2/2z$ only at a distance from 
the jellium edge of several Fermi wavelengths ($\sim 8 \lambda_F$). We remind here that point {\it (i)}
above was our main assumption in the derivation of the slab asymptotic limit of Eq.~(\ref{sdpasymp2}).
This assumption is fully justified after the numerical results shown in Fig.~3.

At this point, with a few algebraic manipulations, we rewrite the asymptote 
of Eq.~(\ref{sdpasymp2}) in the physically motivated image-like form
\begin{equation}
\varepsilon_{x}^{\text{Slab}}(z\rightarrow \infty) \rightarrow - \; \alpha_{x}^{\text{Slab}} \frac{ e^{2}}
{\left[z-z_{x}^{\text{Slab}}(d,r_{s},z)\right]},    
\label{image.like}
\end{equation}
where
\begin{equation}
\alpha_{x}^{\text{Slab}} = 1/2,
\label{alpha.x}
\end{equation} 
and $z_{x}^{\text{Slab}}(d,r_{s},z)$, which represents the location of the so-called image plane, results in 
\begin{equation}
z_{x}^{\text{Slab}}(d,r_{s},z)= \beta_{m}(d,r_{s})+\frac{\gamma_{m}(d,r_{s}) - [\beta_{m}(d,r_{s})]^2}{z} + O \left( \frac{1}{z^{2}} \right).
\label{zeta.cero}
\end{equation}
As $z \rightarrow \infty$, $z_{x}^{\text{Slab}}(d,r_{s},z)$ reaches a finite value, given by
\begin{equation}
z_{x}^{\text{Slab}}(d,r_{s},z \rightarrow \infty) \rightarrow \beta_{m}(d,r_{s}) = \overline{z}^m(d,r_{s}) -  2 /
\pi k_F^m(d,r_{s})= -d/2 -2 / \pi k_F^m(d,r_{s}).\label{zeta}
\end{equation}

\begin{figure}
\includegraphics*[scale=0.5,angle=-90]{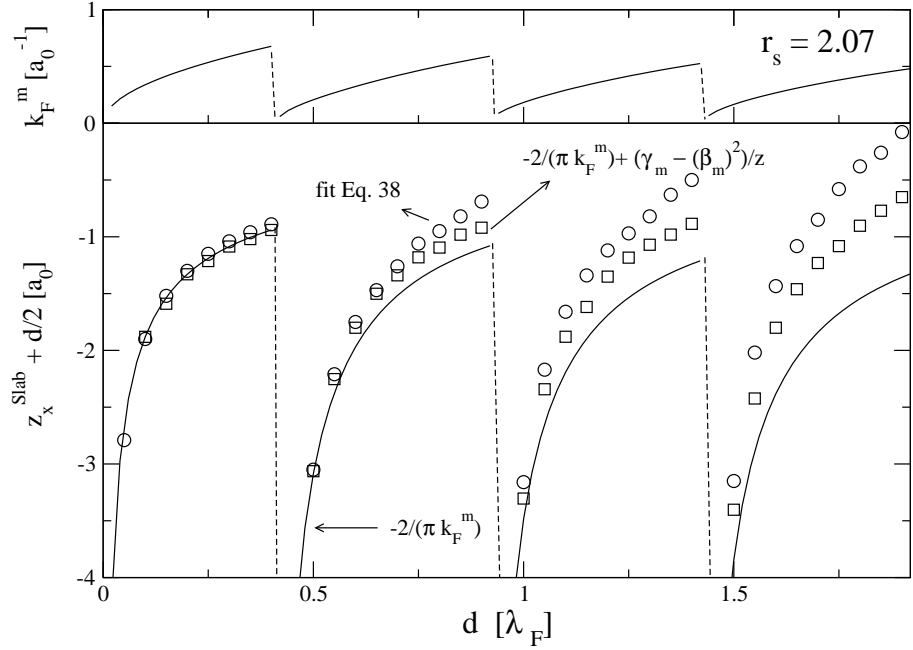}
\caption{Parameter $z_{x}^{\text{Slab}}(d,r_{s}=2.07,z)$ from Eq.~(\ref{image.like}), 
as a function of $d$, in several
approximations. Full line, $z_{x}^{\text{Slab}}(d,r_{s}=2.07,z\rightarrow \infty) +d/2 = - 2 / \pi k_{F}^{m}(d,r_{s}=2.07)$; 
squares, Eq.~(\ref{zeta.cero});
circles, fit to Eq.~(\ref{image.like}). Upper panel, $k_{F}^{m}(d)$ as a function of $d$.}
\label{f4}
\end{figure}

In Fig.~4, we plot a comparison of Eq.~(\ref{zeta}) (solid line) with the image-plane 
position that we obtain by fitting our full numerical calculation of Eq.~(\ref{5}) with 
the image-like Eq.~(\ref{image.like}) and $\alpha_{x}^{\text{Slab}} = 1/2$ (empty circles).
For this, we have used a fit region of width $2 \lambda_{F}$ centered at $ 6 \lambda_{F}$
from the jellium edge in the vacuum.  
Differences between Eq.~(\ref{zeta}) (solid line) and our numerical estimate (empty circles), 
which in the case of very thin films are negligible, are entirely due to the fact that the 
fitting of the numerical calculation must be carried out in a vacuum region that extends very far away 
from the surface. Empty squares correspond to the
result of Eq.~(\ref{zeta.cero}), including the correction to the leading
term, and taking $z = 6 \lambda_{F}$ (which is the average value of the fit region indicated above). 
According to Eq.~(\ref{zeta}), $z_{x}^{\text{Slab}}(d)+d/2$ is inversely proportional to
$k_F^m(d)$, which exhibits an oscillatory behavior as a function of the slab width $d$ (see the solid line in the upper part of Fig.~4) going to zero every time a SDL becomes occupied. Hence, the location of the image plane becomes infinitely negative every time a SDL becomes occupied, which results in the strong finite-size oscillations shown in Fig.~4.

\subsection{Semi-infinite jellium}

\begin{figure}
\includegraphics*[scale=0.5,angle=-90]{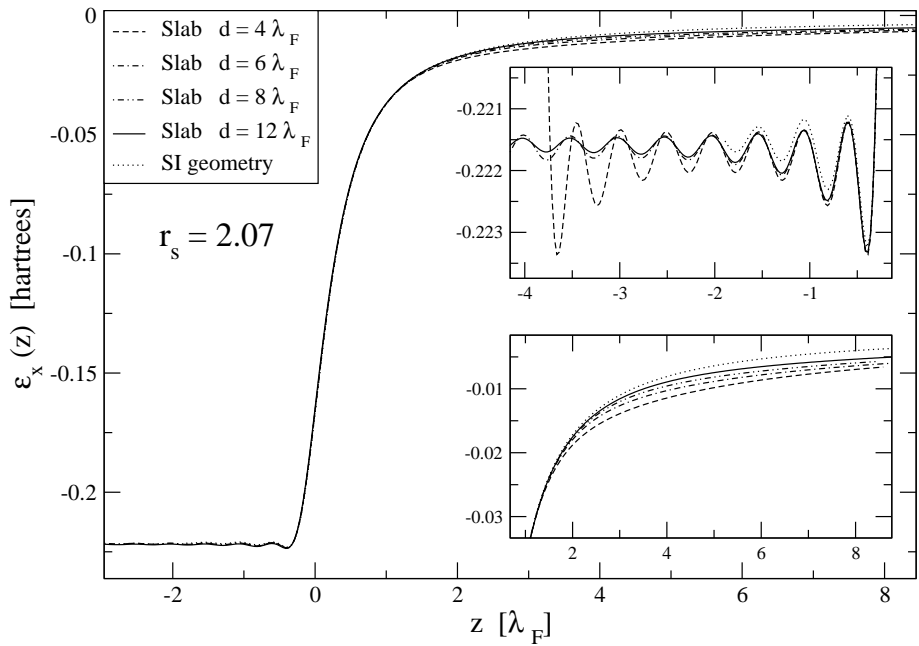}
\caption{$\varepsilon_{x}^{\text{Slab}}(z)$ for $r_s = 2.07$, 
slabs with $d = 4, 6, 8$, and $12 \; \lambda_F$. 
Dotted line, $\varepsilon_{x}^{\text{SI}}(z)$ for the semi-infinite case, from Eq.~(\ref{5.semi}). 
Upper inset: enlarged view of the bulk region. Lower inset: asymptotic region.}
\label{f5}
\end{figure}

Now we focus on a comparison between our full numerical jellium-slab and semi-infinite-jellium 
calculations of the position-dependent exchange energy per particle (see Fig.~5). 
In the bulk, as $d$ increases the slab calculations converge with the semi-infinite calculation 
(see the inset at the upper part of Fig.~5), and both slab and semi-infinite 
calculations approach in the bulk region far away from the surface the exchange 
energy per particle of a three-dimensional (3D) homogeneous 
electron gas, $\varepsilon_x^{\text{3D}}/(e^2/a_0) = -(3 / 4 \pi) (9 \pi /4)^{1/3}/r_s \approx - \; 0.22134 $. 
In the vacuum, however, there is always a region far enough away from the surface where the 
jellium slab and the semi-infinite jellium behave differently: while all slab calculations 
converge to an image-like behavior of the form of Eq.~(\ref{image.like}) with
$\alpha_{x}^{\text{Slab}} = 1/2$, the semi-infinite
$\varepsilon _{x}^{\text{SI}}(z\rightarrow\infty)$ exhibits an image-like behavior in agreement  
with Eq.~(\ref{semi}) (see Fig.~6). 
Fig.~5 also shows that as the width $d$ increases the slab $\varepsilon _{x}^{\text{Slab}}(z)$
coincides with the semi-infinite $\varepsilon _{x}^{\text{SI}}(z)$ in a wider vacuum region near
the surface (see the lower inset of Fig.~5).

\begin{figure}
\includegraphics*[scale=0.5,angle=-90]{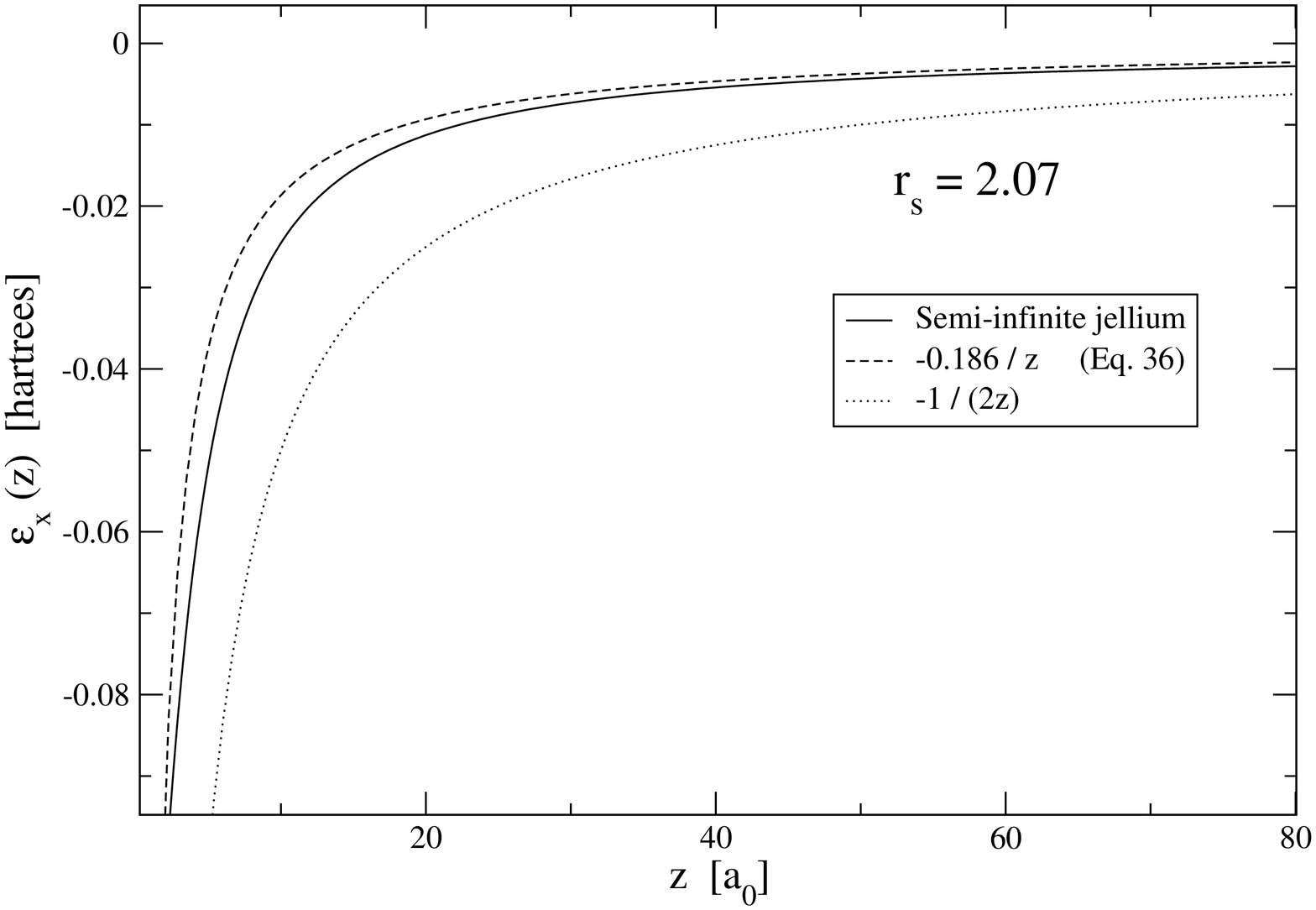}
\caption{Asymptotic behavior of $\varepsilon_{x}^{\text{SI}  }(z)$ for the semi-infinite 
case, and comparison with Eq.~(\ref{semi}) and the asymptote form of
Eq.~(\ref{sdpasymp2}).}
\label{f6}
\end{figure}

Fig.~6 shows a comparison of our full numerical calculation 
of Eq.~(\ref{5.semi}) with the asymptote of Eq.~(\ref{semi})
(solid and dashed lines, respectively) for $r_s=2.07$.
In this case, $\alpha=\sqrt{\varepsilon_F/W}=2.048$ (as obtained from our exchange-only 
LDA self-consistent calculation of the work function $W$) and Eq.~(\ref{semi}) yields
$\varepsilon _{x}^{\text{SI}}(z\rightarrow \infty ) \rightarrow -0.18622 \; e^2/z$ (dashed line),  
which is in contrast with the asymptote of
Eq.~(\ref{sdpasymp2}) (dotted line) that holds in the case of jellium slabs.
The same comparison has been done for other values of $r_s$, and we have found 
that our full numerical calculation is always very close (as in Fig.~6) to the 
asymptote of Eq.~(\ref{semi}).

\begin{figure}
\includegraphics*[scale=0.5,angle=-90]{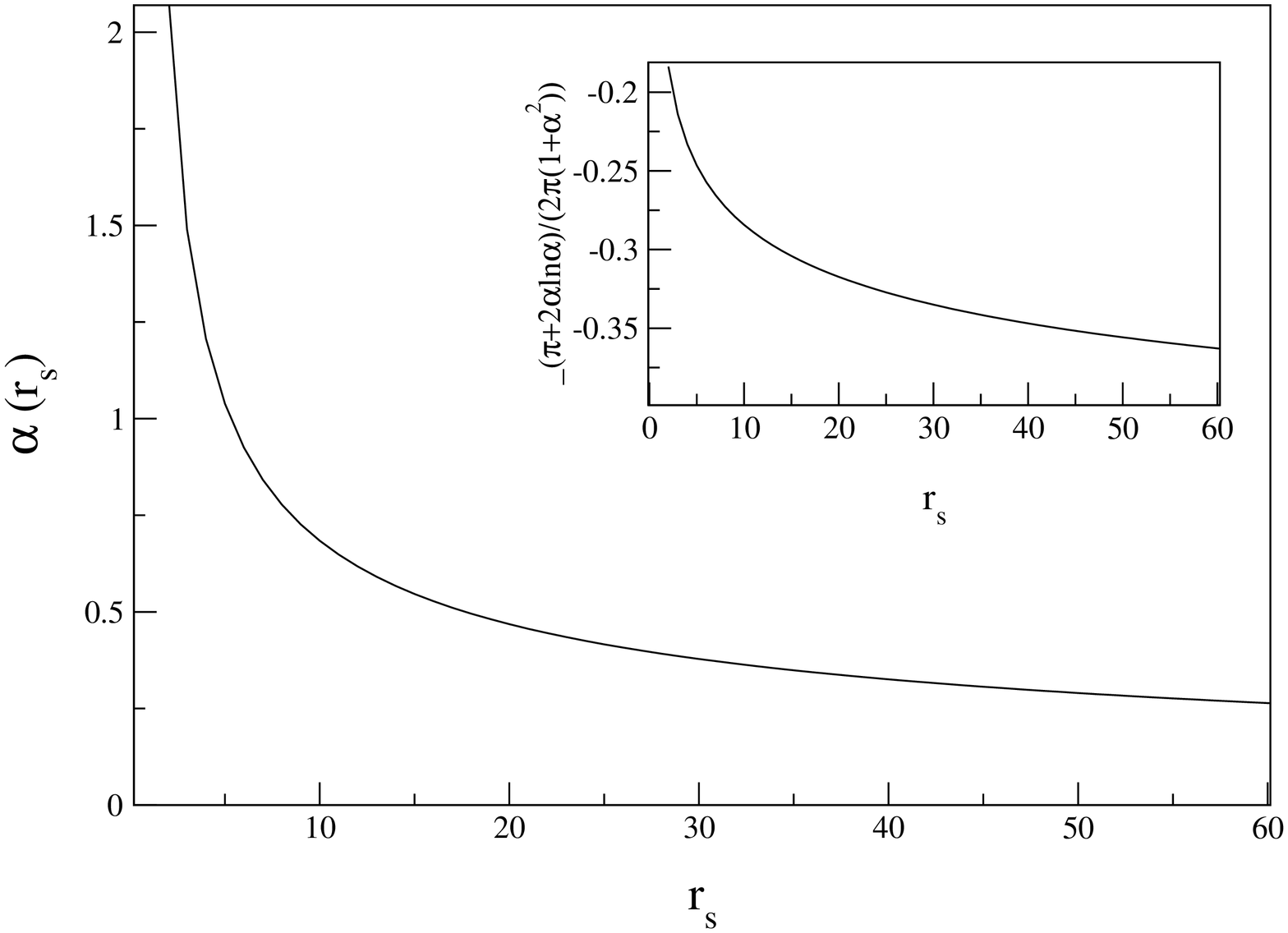}
\caption{$\alpha(r_s) = \sqrt{\varepsilon_{F}(r_s) / W(r_s)}$ versus $r_{s}$, for the
semi-infinite case. Inset, the coefficient 
$-\left[\pi + 2 \alpha \ln (\alpha)\right]/\left[2 \pi (1+\alpha^2)\right]$ of Eq.~(\ref{semi}) versus $r_{s}$.}
\label{f7}
\end{figure}

Finally, we display in Fig.~7 our exchange-only LDA self-consistent calculation of the 
coefficient $\alpha = \sqrt{\varepsilon_F/W}$ in a wide range of electron densities.
It is important to note that the corresponding coefficient 
$\left[\pi + 2 \alpha \ln (\alpha)\right]/\left[2 \pi (1+\alpha^2)\right]$ (see the inset to Fig.~7) entering 
Eq.~(\ref{semi}) is close to $1/4$ at metallic densities ($r_s=2-6$). Fig.~7 also shows that only at extremely low densities the coefficient $\alpha$  
approaches zero, thereby the coefficient $\left[\pi + 2 \alpha \ln (\alpha)\right]/\left[2 \pi (1+\alpha^2)\right]$ of 
Eq.~(\ref{semi})  approaching $1/2$. Hence, the asymptotic limits of $\varepsilon _{x}^{\text{Slab}}(z)$
and  $\varepsilon _{x}^{\text{SI}}(z)$ only coincide in the low-density limit ($r_s\to\infty$).

\section{Summary and conclusions}

We have presented a detailed analysis of the position-dependent exchange energy per particle
$\varepsilon_x(z)$ at jellium slabs and the semi-infinite jellium.

For jellium slabs, we have found that in the vacuum region far away from the surface
$\varepsilon_x^{\text{Slab}}(z \rightarrow \infty) \rightarrow - \; e^2/(2z)$,
independent of the bulk electron density. This is the equivalent to the well-known result
$\varepsilon_x({\bf r} \rightarrow \infty) \rightarrow - \; e^2/(2r)$, which holds in the case of localized
finite systems like atoms and molecules.~\cite{dreizler}
The equivalence between these results is however not straightforward, since slabs have an extended character
in the $x-y$ plane, being ``localized'' only along the
$z$-coordinate. In the vacuum side of the surface there is a region where 
$\varepsilon_x^{\text{Slab}}(z)$ 
coincides with $\varepsilon_x^{\text{SI}}(z)$ and this region increases as $d$ increases.
The fitting of our numerical calculations of
$\varepsilon_x^{\text{Slab}}(z)$ to a physically motivated image-like expression is
feasible, but the resulting location of the image plane 
$[z_x^{\text{Slab}}(d,r_{s},z)]$ shows strong finite-size oscillations.
In particular, we have shown analytically that $z_x^{\text{Slab}}(d,r_{s},z\rightarrow \infty) 
= -d/2 - 2 / \pi k_F^m(d,r_{s})$, $k_F^m(d,r_{s})$ being a signature of
the energy of the highest occupied SDL with respect to the Fermi level.

For a semi-infinite jellium, we have found that our numerical 
calculations agree well with the analytical asymptote [see Eq.~(\ref{semi})] obtained
in Refs.~\onlinecite{ss,ssb,nastos} and \onlinecite{qs}, which approaches the slab asymptote
$- \; e^{2}/2z$ only in the extreme low-density limit ($r_s\to\infty$).    

We attribute the qualitatively different behavior
of $\varepsilon_x^{\text{Slab}}(z \rightarrow \infty)$
and $\varepsilon_x^{\text{SI}}(z \rightarrow \infty)$ to the fact that 
these asymptotes
are approached in different ranges. While in the case of the semi-infinite jellium the 
asymptote is reached at
distances $z$ from the surface that are large compared to the Fermi wavelength 
(the only existing
length scale in this model), for slabs the asymptote is reached at distances $z$ from 
the surface that are
large compared to $1/k_F^m$ (which is typically larger than the slab thickness $d$). For thick slabs with $d>>\lambda_F$ ($\lambda_F$ being the Fermi wavelength), $\varepsilon_x^{\text{Slab}}(z)$ first coincides with $\varepsilon_x^{\text{SI}}(z)$ [dictated by Eq.~(36) at $z>>\lambda_F$] in the vacuum region near the surface (see Fig.~5), but at distances from the surface that are large compared to $1/k_F^m$, $\varepsilon_x^{\text{Slab}}(z)$ turns to the slab image-like behavior of the form of Eq.~(22) [or, equivalently, Eq.~(38) with $\alpha_x^{\text{Slab}}=1/2$]; in the limit as $d\to\infty$ (i.e., when the jellium slab becomes semi-infinite), $\varepsilon_x^{\text{Slab}}(z\to\infty)$ {\it coincides} with $\varepsilon_x^{\text{SI}}(z\to\infty)$ {\it everywhere}. In the low-density limit, where $\lambda_F\to\infty$, the condition $d>>\lambda_F$ is never fulfilled and $\varepsilon_x^{\text{Slab}}(z)$ reaches (at $z>>1/k_F^m$) one {\it single} asymptote: the slab image-like behavior of the form of Eq.~(22) [or, equivalently, Eq.~(38) with $\alpha_x^{\text{Slab}}=1/2$], which turns out to coincide with the semi-infinite-jellium asymptotic behavior dictated by Eq.~(36).
      
Finally, we note that as $\varepsilon_{xc}(z) = \varepsilon_x(z) + \varepsilon_c(z)$, the same conclusion 
is expected to be valid for the exchange contribution to the position-dependent {\it xc} energy per particle. Recent developments concerning the asymptotic behavior of the correlation contribution to the KS exchange correlation potential $V_{xc}(z)$ of a semi-infinite jellium can be found in Ref.~\onlinecite{qs}.  

\section{Acknowledgments}

C.M.H wishes to acknowledge the financial support received
from CONICET of Argentina. C.R.P. was supported by the European Community 
through a Marie Curie IIF (MIF1-CT-2006-040222) and CONICET of Argentina
through PIP 5254. J.M.P. acknowledges
partial support by the University of the Basque Country, the Basque
Unibertsitate eta Ikerketa Saila, and the Spanish Ministerio de Educaci\'on y
Ciencia (Grants No. FIS2006-01343 and CSD2006-53). 

\begin{appendix}
\section{Derivation of asymptotic expressions}
Here we will shown in detail how to go through Eqs.~(\ref{7})-(\ref{sdpasymp}) in the text.
Starting from Eq.~(\ref{7}), one first notice that\cite{Mathematica}
\begin{equation}\label{A1}
\int\limits_{0}^{\infty} \frac{dx}{x} \frac{J_1^2(ax)}{\sqrt{x^2+y^2}} = \frac{1}{2|y|}
\left[1-\frac{I_1(2a|y|)}{a|y|}+\frac{L_1(2a|y|)}{a|y|} \right] \; ,
\end{equation}
with $I_1$ and $L_1$ being the modified Bessel and Struve functions, respectively. Substitution
of Eq.~(\ref{A1}) in Eq.~(\ref{7}) yields at once Eq.~(\ref{8}). Now, in the asymptotic limit,\cite{AS1}
\begin{equation}\label{A2}
L_1(x \gg 1) \rightarrow I_1(x \gg 1) - \frac{2}{\pi} + \frac{2}{x^2} - ... \; ,
\end{equation}
which inserted in the definition of the function $F(x)$ yields Eq.~(\ref{9}). Substitution of this
expansion for $F(x)$ in Eq.~(\ref{8}) leads to the expression,
\begin{eqnarray}
\varepsilon_{x,1}^{\text{Slab}}(z \rightarrow \infty) \rightarrow - \frac{e^2}{2} \int\limits_{- \infty}^{\infty}
dz^{\prime}\frac {|\xi_1(z^{\prime})|^2}{|z-z^{\prime}|}\left[1-\frac{2}{\pi}\frac{1}{k_F^1|z-z^{\prime}|}+
\frac{1}{2}\frac{1}{(k_F^1|z-z^{\prime}|)^3} -... \right] \; , \\
= -\frac{e^2}{2} \int\limits_{-\infty}^{\infty}dz^{\prime}\frac{|\xi_1(z^{\prime})|^2}{|z-z^{\prime}|} + 
\frac{e^2}{\pi k_F^1}\int\limits_{-\infty}^{\infty}dz^{\prime}\frac{|\xi_1(z^{\prime})|^2}{|z-z^{\prime}|^2}
-\frac{e^2}{4(k_F^1)^3}\int\limits_{-\infty}^{\infty}dz^{\prime}\frac{|\xi_1(z^{\prime})|^2}{|z-z^{\prime}|^4} + ... \;. \label{A4}
\end{eqnarray}
Expanding the denominators of Eq.~(\ref{A4}) in the large $z$ limit, the different contributions
in Eq.~(\ref{sdpasymp}) arise. For instance, the leading contribution $-e^2/(2z)$ comes from the first
term on the r.h.s. in Eq.~(\ref{A4}), with $|z-z^{\prime}|$ approximated by $z$. It is interesting to note that one important
feature of this result, as is its material and slab-size independence, is consequence (in this context) 
of the normalization of the
SDL wave-functions. On more general grounds, and returning to the alternative definition of $\varepsilon_x^{\text{Slab}}(z)$
given in Eq.~(\ref{physical}), this is more physically understood as a consequence of that the integral
of $h_x(z;\rho,z+Z)$ over all possible ``observational'' coordinates ($\bm{\rho},Z$) is exactly $-1$.\cite{rrp}
The next term in the expansion, proportional to $\beta^1$ and with decay $z^{-2}$, is obtained from the
sub-leading contribution of the first term on the r.h.s. in Eq.~(\ref{A4}), together with the leading contribution
from the second term. To the order explicitly displayed in Eq.~(\ref{sdpasymp}), no contribution arises
from the last (third) term in Eq.~(\ref{A4}), as the leading contribution coming from this term 
to $\varepsilon_{x,1}^{\text{Slab}}(z \rightarrow \infty )$ is of the order $z^{-4}$. While this analysis
has been performed for the single-occupied SDL case, it also applies to the general case where more
than a SDL is occupied, as explained in Section II.A.2. 

\end{appendix}


\begin{thebibliography}{[Vo]}
\bibitem{note0} $\bar n$ is given by $\bar n = 3/(4 \pi r_s^3 a_0^3)$,
with $r_s$ being the dimensionless electron-density parameter defined as the radius
of a sphere containing on average one electron and $a_0$ being the Bohr radius. 
A convenient length unit for the present system is the Fermi wavelength
$\lambda _{F}=(32\pi ^{2}/9)^{1/3}\;r_{s}a_{0} \approx 3.274\; r_{s}a_{0} $.
\bibitem{hartree.a} This Hartree potential includes the contribution coming from the uniform positive background (proportional to $n_{+}$). Alternatively, this contribution could have been  denoted separately as the "external potential".
\bibitem{dens.aa} The dimensions of Eq.~(\ref{density}) are (length)$^{-3}$, as corresponds 
to our 3D system. However, because of the slab geometry the number density here 
only depends on one spatial coordinate $z$.
\bibitem{oneSDL} As $(k_{F}^{1})^{2} = 2 m_e (\varepsilon_F - \varepsilon_1) / \hbar^{2}  = 2 \pi d \overline{n}$,
this limit $(\varepsilon_1 < \varepsilon_F < \varepsilon_2)$ can be achieved either for narrow slabs ($d \rightarrow 0$) or in the low-density limit $\overline{n}\rightarrow 0$.
\bibitem{note.bb} Note that $\overline{z}^m =-d/2$ and $\overline{z^2}^m >0$. Thus,
$\beta_m$ is always negative as a sum of two negative terms, and $\gamma_m$ 
is always positive as a sum of two positive terms.
\bibitem{notesahni} The definition of $W$ entering Eq.~(23) of Ref.~\onlinecite{ssb} is not the work function (as defined here), but the barrier height ($W^{SS}=W+\varepsilon_F$) instead.
\bibitem{note0a} Most of the calculations presented here have been found to be well converged 
by locating the two infinite barriers at $8 \lambda _{F}$ from each
jellium edge.  For the case of one single occupied SDL, the two infinite barriers have been taken
at $10 \lambda_{F}$ from each jellium edge.
\end{thebibliography}
\end{document}